\documentstyle[12pt]{article}
\begin{document}
\baselineskip 12pt
\begin{titlepage}
\title{On the thermal stability of transonic accretion discs} 
\author{Ewa Szuszkiewicz$^{1,2}$ and John C. Miller$^{2,3}$}
\date {}
\maketitle
\protect{\thispagestyle{empty}}
\begin{center}
$^1$Astronomy Group, Department of Physics and Astronomy, \\
University of Leicester, University Road, \\ 
Leicester $\,$LE1 7RH, England \\
\end{center}
\begin{center}
$^2$International School for Advanced Studies, SISSA,\\
via Beirut 2-4, I-34013 Trieste, Italy \\
\end{center}
\begin{center}
$^3$Nuclear and Astrophysics Laboratory, University of Oxford, \\
Keble Road, Oxford $\,$OX1 3RH, England  \\
\end{center}
\vspace{3.0cm}
\vspace{2.0cm}
\begin{center}
Accepted for publication in the MNRAS (January 1997)
\end{center}
\end{titlepage}
\begin{center}
{\bf Abstract} 
\end{center}

Nonlinear time-dependent calculations have been carried out in order to
study the evolution of the thermal instability for vertically
integrated, non self-gravitating models of optically thick, transonic,
slim accretion discs around black holes.  In these calculations we
investigated only the original version of the slim disc model with low
viscosity ($\alpha = 0.001$) and for a stellar mass ($10\, M_{\odot}$)
black hole.  This original version of the model does not yet include
several important non-local effects (viscous forces in the radial
equation of motion, diffusion-type formulation for the viscosity in the
angular momentum equation, viscous dissipation rate associated with the
stress in the azimuthal direction and radiative losses in the radial
direction in the energy balance equation). It is clear, therefore, that
this treatment is greatly simplified but our strategy is to consider
this as a standard reference against which to compare results from
forthcoming studies in which the additional effects will be added one
by one thus giving a systematic way of understanding the contribution
from each of them.
 
We have considered an already-formed disc and studied its stability against
small axisymmetric perturbations. Those models which were stable according
to local analysis, remain stable and stationary to a good approximation as
also do models for which local analysis predicted an unstable region with
radial dimension smaller than the shortest wavelength of the unstable
modes. In terms of luminosity, all models with luminosity less than or
equal to $0.08L_{_E}$ are stable. For models with higher luminosity than
this but which are still sub-Eddington, a shock-like structure
forms near to the sonic point, probably leading to subsequent disruption
of the disc. The model with Eddington luminosity evolves in a violent way,
with the shock-like structure being formed already within the first second
of evolution. From an examination of phase space trajectories, our
preliminary conclusion is that the stabilizing effect of advection is not
strong enough in these models to allow for limit cycle behaviour to occur.
However, in order to make a definitive statement on this, it would be
necessary to implement special numerical techniques for treatment of the
shock-like structure, which becomes very extreme, and this lies beyond the
scope of the present paper. 

{\it Subject headings}: accretion: accretion discs, instabilities: thermal  
\newpage
\section {Introduction}

Accretion discs around black holes are thought to be present in galactic
black hole candidates and active galactic nuclei. One of the
characteristic properties of these objects is their variability. Apart
from fundamental questions concerning the very existence of fluid
disc-like configurations and their lifetimes, the most immediate purpose
of the study of accretion disc stability is to discover the origin of the
incipient instabilities which may be responsible for the observed
variability.  However, before applying the results of stability studies to
realistic objects, one must eliminate the possibility that the behaviour
of the model is merely a consequence of the approximations made in its
construction (Piran, 1978; Pringle, 1981). 

Equilibrium models of geometrically {\it thin} accretion discs were found
to be subject to two types of instability -- thermal (Pringle, Rees, and
Pacholczyk, 1973) and viscous (Lightman and Eardley, 1974). A thorough
investigation of these was carried out by Shakura and Sunyaev (1976) who
found that the instabilities develop via a bifurcation of unstable modes. 
Both types of instability occur when radiation provides a substantial
contribution to the total pressure. In thin discs, the thermal time scales
are generally much shorter than the viscous ones and so thermal
instabilities, if present, dominate over viscous instabilities. The first
discussions of this were made in terms of local linear analysis, in which
the global response of the disc and nonlinear effects were not taken into
account, but then Lightman (1974) carried out time-dependent calculations
to examine the global behaviour.  Detailed time-dependent disc evolution
was studied first in the context of dwarf novae (Meyer and Meyer-Hofmeister,
1982; Smak, 1982;
Papaloizou, Faulkner and
Lin, 1983). These authors 
 showed that the dwarf nova outbursts arise from collective
relaxation oscillations of accretion discs in which cool hydrogen ionization
regions provide a triggering mechanism.  As a guide for
understanding the global instabilities, an associated local analysis was
carried out and a fundamental role in this was played by the fact that
when the values of the accretion rate, $\dot M$, and the surface density,
$\Sigma$, at a particular location (value of $r$) are plotted against each
other for a sequence of equilibrium models, the points form characteristic
S-shaped curves in the $\log \dot M - \log \Sigma$ plane.  Analogous
investigations of the time-dependent evolution of accretion discs around
neutron stars and black holes were started by Taam and Lin (1984) who
found that if the viscous stress tensor is proportional to the total pressure,
global analysis confirms the results inferred from local analysis and the
instability manifests itself in terms of short time-scale luminosity
fluctuations. Similar studies were performed subsequently by Lasota and
Pelat (1991) and more recently by Chen and Taam (1994). Lasota and Pelat
concluded that the (locally) thermally unstable region in the inner part
of the disc always becomes effectively optically thin and that the
evolution and disc structure depends sensitively on the treatment of the
transitions between the optically thick and optically thin regimes.  Chen
and Taam (1994) investigated the global evolution of thermal-viscous
instabilities in the case where the viscous stress is proportional to the
gas pressure with a particular form for the viscosity parameter.  Their
calculations show that the instabilities are globally coherent throughout
the entire unstable region of the disc. 

All of the authors mentioned above studied Keplerian discs.  The only
investigation which we know of in which a full set of time-dependent
equations has been solved, is the one by Honma, Matsumoto and Kato (1991). 

The importance for accretion discs around black holes of the transonic
nature of the flow, deviations from Keplerian rotation and non-local
cooling by advection, was first pointed out by Paczy\'{n}ski and
Bisnovatyi-Kogan (1981) who constructed a simple global equilibrium model
of a geometrically thin steady disc. 
A full treatment, 
including all of these effects, 
was formulated by Loska (1982) and
Muchotrzeb and Paczy\'nski (1982).
Following their approach, Abramowicz, Czerny, Lasota and Szuszkiewicz
(1988) calculated a sequence of slim accretion disc models for a wide
range of accretion rates and found S-shaped curves in $\log \dot M - \log
\Sigma$ plane, a result which was later confirmed by Abramowicz, Kato and
Matsumoto (1989).  This finding recalls the situation for dwarf novae
(although the physical processes leading to this shape are quite different
in the two cases) and this encouraged Honma, Matsumoto and Kato (1991) to
study the nonlinear evolution of the thermally unstable models to see
whether similar behaviour occurred. They found that, at least for high
viscosities ($\alpha =0.1$), a limit-cycle behaviour does indeed occur
which is very reminiscent of that for dwarf novae. 

The aim of our present study was to examine the global behaviour of the
thermally unstable slim disc models in their original formulation
(Abramowicz, Czerny, Lasota and Szuszkiewicz, 1988) with low viscosity
($\alpha=0.001$). The effects of advective cooling are included and our
results represent a suitable standard reference for making comparison with
results from forthcoming calculations in which further non-local processes
will be included which were not present in the original version of the
slim disc model. 

In Section 2, we discuss the set of basic equations used for our
calculations, pointing out which assumptions are made and describing the
averaging procedure employed. The numerical treatment of the equations is
presented in Section 3. In Section 4 the global behaviour of the models is
shown and in Section 5 the same models are viewed locally. Section 6
contains comments and conclusions. 

\section {Basic equations}
In this section, we present the equations used for calculating 
non-stationary behaviour of the original slim disc model and highlight 
the assumptions involved in this treatment.

\subsection {Evolution equations}

We consider here an axisymmetric non-self-gravitating fluid disc formed in
the gravitational potential $\Phi$ of a black hole with mass $M$ and use
cylindrical polar coordinates ($r$, $\varphi$, $z$) centred on the black
hole.  The symmetry axis, $z$, coincides with the rotation axis of the
disc. The behaviour of the fluid is governed by the basic hydrodynamical
equations for conservation of mass, energy and momentum (radial, angular
and vertical). We study a one-dimensional formulation in which most of the
equations are written in a vertically integrated form with the vertical
structure being approximated under the assumptions that hydrostatic
balance always holds in the vertical direction and that flow properties
above and below the equator ($z=0$) are the same. In this picture, the
velocity field is approximated by $v_z=0$, $v_r=v_r(r,t)$ and $v_{\varphi}
= v_{\varphi}(r,t)$. Newtonian mechanics is used and general relativistic
effects, which become significant near to the inner edge of the disc, are
simulated by using the pseudo-Newtonian potential introduced by
Paczy\'nski and Wiita (1980): 

\begin{equation}
\Phi = - {GM \over \sqrt{r^2+z^2} -r_{_G}}.
\end{equation}

Here $r_{_G}$ is the Schwarzschild radius defined by $r_{_G}= 2GM/c^2$, where
$G$ and $c$ are the gravitational constant and velocity of light
respectively.

Within this framework, the evolution equations can be written in the 
following form:

{\it -- Mass conservation equation:}

\begin{equation}
{D \Sigma \over D t} = - {\Sigma \over r} {\partial \over \partial r}
\left( r v_r \right)
\end{equation}

where $D/Dt$ is the Lagrangian derivative given by 

$$ {D \over Dt} ={\partial \over \partial t} 
+ v_r {\partial \over \partial r },
$$

$\Sigma=
\Sigma(r,t)$ is the surface density obtained by vertically integrating the 
volume density $\rho$
\begin{equation}
\Sigma = 2 \int_{0}^{H} \rho \, dz
\end{equation}
with $H$ being the half-thickness of the disc, and $v_r = Dr/Dt $ (which is
negative for an inflow). For the stationary case, the mass conservation
equation reduces to the statement that the local accretion rate is the
same at all locations in the disc
\begin{equation}
\dot m (r) = -2\pi r \Sigma v_r = \dot M ,
\end{equation} 

where $m(r)$ is the total mass internal to the cylindrical radius $r$, 
$\dot m = \partial m / \partial t $ and $\dot M $ is the rate of increase 
of mass of the black hole.

{\it -- Radial equation of motion:}
\begin{equation}
{Dv_r \over Dt} =
- {1 \over \rho  }{\partial p \over \partial r} +
 {{\left( l^2 - l_{_K}^2 \right) } \over r^3} + F_{rr}  
 - F_{\varphi\varphi},
\end{equation}
where $p$ is the pressure, $l=l(r,t)=rv_{\varphi}(r,t)$ is the specific
angular momentum, $l_{_K}$ is the value of $l$ for Keplerian motion 
with $v_{\varphi}=  
\left[-r \left( d\Phi/dr\right)\right]^{1/2} 
= \left[ GMr/(r-r_{_G})^2 \right]^{1/2}$.
$F_{rr}$ and $F_{\varphi \varphi}$ are the viscous
forces: 
$$ F_{rr} = {1 \over r\rho  } {\partial \over \partial r}
\left( r \tau_{rr} \right)$$
$$ F_{\varphi \varphi} = {1 \over r\rho  } 
\left( \tau_{\varphi \varphi}  \right)$$
where
$$\tau_{rr} = {2 \over 3} \mu \left[ 2{\partial v_r \over \partial r}
- {v_r \over r} \right]$$
$$\tau_{\varphi \varphi} = {2 \over 3} \mu \left[ {2 v_r \over  r}
- {\partial v_r \over \partial r} \right]$$
and $\mu$ is the shear viscosity coefficient (bulk and radiative
viscosities are not considered anywhere in the present treatment). The 
importance of the viscous forces $F_{rr}$ and $F_{\varphi \varphi}$ has 
been pointed out (e.g. by Papaloizou and Stanley, 1986) and they have  
been considered in the more recent studies but we do not include them 
here since they were not part of the original slim disc model. 
Also, in accordance with the original slim disc model, 
we do not vertically integrate the radial equation of motion
but, instead, solve it in the equatorial plane.
We comment further on this in Section 2.2.

The stationary version of Eq.~(5) is 
\begin{equation}
v_r {dv_r \over dr} =
- {1 \over \rho  }{dp \over dr} +
{{\left( l^2 - l_{_K}^2 \right) } \over r^3}
\end{equation}

{\it -- Azimuthal equation of motion, for the specific angular momentum, 
$l$:}

\begin{equation}
{Dl \over Dt}=
{1 \over {r \Sigma}} {\partial \over \partial r} \left( r^2
2\int_0^H\tau_{\varphi r}dz\right)
\end{equation}

where the rate of shearing is 
\begin{equation}
\tau_{\varphi r} =  \mu r
{\partial \Omega \over \partial r}
\end{equation}
which, following Shakura and Sunyaev (1973), is often approximated by 
\begin{equation}
\tau_{\varphi r} = - \alpha p 
\end{equation}
where $\alpha$ is a (constant) viscosity parameter. This form, which can
be derived from Eq.~(8) in the case of a (stationary) Newtonian Keplerian
disc and with the aid of several additional approximations, is the one
used in the present paper in accordance with the original slim disc model.
However, a major objective of our future work concerns investigation of
the extent to which inclusion of a more physical viscosity prescription
causes differences in non-stationary behaviour with respect to that of the
``standard'' models calculated here. With expression (9), Eq.~(7) can be 
rewritten as

\begin{equation}
{Dl \over Dt}= -
{\alpha \over {r \Sigma}} {\partial \over \partial r} \left( r^2 P
\right)
\end{equation}

where $P$ is the vertically-integrated pressure

\begin{equation}
P = 2 \int_{0}^{H} p \, dz
\end{equation}

For a stationary model, Eq.~(10) can be integrated explicitly and, using
the boundary condition that there is no viscous torque at the black hole
horizon, this gives
\begin{equation}
\dot M(l-l_0) = 2\pi \alpha P r^2
\end{equation}
where $l_0$ is the specific angular momentum at the inner edge of the disc. 

{\it -- Vertical equation of motion:}

\begin{equation}
{D v_z \over Dt} = - 
{1 \over \rho }{\partial p \over \partial z} - {\partial \Phi \over 
\partial z} + F_{zz},
\end{equation}
where the the viscous force $F_{zz}$ is given by
$$F_{zz} = {1 \over \rho} {\partial  \over \partial z} \tau_{zz}
$$ 
with 
$$\tau_{zz} = -{2 \mu \over 3r} {\partial \over \partial r}(rv_r) .
$$
However, $F_{zz}$ is not included in the present treatment. Also, since we
are considering hydrostatic equilibrium in the vertical direction, the
acceleration term vanishes. We note that neglecting the vertical
acceleration may be a rather poor approximation in some parts of the disc,
as suggested by two-dimensional studies (Papaloizou and Szuszkiewicz,
1994), and investigation of the effect of {\it not} neglecting it is one
of the important issues to be addressed in our forthcoming study.  The
approach used here is exactly the same as for  the stationary slim disc
models and is discussed further in Section 2.2. 

{\it -- Conservation of thermal energy:}

\begin{equation}
\rho T {DS \over Dt} = Q_{vis} + Q_{rad} ,
\end{equation}
where $S$ is the entropy per unit mass, $T$ is the temperature, $Q_{vis}$
is the rate at which heat is generated by viscous friction (including
the effects of stresses in both the azimuthal and radial directions)
\begin{equation}
Q_{vis} = {4 \over 3}\mu \left[ \left( {\partial v_r \over 
\partial r} \right)^2+
\left( { v_r \over r} \right)^2
- {v_r \over r}{\partial v_r \over \partial r}\right] +
\mu \left(r{\partial \Omega \over \partial r}\right)^2 ,
\end{equation}

and $Q_{rad}$ is the rate at which heat is lost or gained by means of
radiative energy transfer (expressed as the sum of terms for the radial
and vertical directions)

\begin{equation}
Q_{rad} = {1 \over r} {\partial \over \partial r} \left[ \chi_{_r} r
{\partial T \over \partial r}\right] + {\partial  \over \partial z}
\left[ \chi_{_r} {\partial T \over \partial z}\right]
\end{equation}

where $\chi_{_r} = 4acT^3/3\kappa \rho$ is the radiative conductivity, $a$
is the radiation constant and $\kappa$ is the opacity. In the present
treatment, the first terms on the right hand side of Eqs. (15) and
(16) are omitted. Thermal conductivity is not considered here.  Using
the strategy discussed in Section 2.2, the vertically-integrated form
of Eq. (14) can be written in the following way

\begin{equation}
{D T \over Dt} = {\alpha r T (\partial \Omega / \partial r) \over
0.67(12-10.5\beta)} - {T F^- \over 0.67pH(12-10.5\beta)} + {(4-3\beta) \over
(12-10.5\beta)}{T \over \rho} {D\rho \over Dt}
\end{equation} 

where $p$, $\rho$ and $T$ are the values of these quantities on the
equatorial plane, $\beta $ is the ratio of the gas pressure to the
total pressure and $F^-$ is the radiative flux away from the disc in
the vertical direction for which we use the expression

$$F^- = {12 a c T^4 \over 3 \kappa \rho H}, 
$$

as discussed in Section 2.2. The stationary form of Eq.~(17) is
 
\begin{equation}
\dot M (l-l_0) (-{d\Omega \over dr}) +0.67 \dot M T {dS \over dr}
=4\pi r F^- .
\end{equation}

The thermodynamic quantities in the equatorial plane are taken to obey the
equation of state

\begin{equation}
p=k\rho T + {a \over 3}T^4
\end{equation}

and the opacity is approximated by the Kramers formula for chemical
abundances corresponding to those of Population I stars

\begin{equation}
\kappa =0.34 \times (1+ 6\times 10^{24}\rho T^{-3.5}) .
\end{equation} 

Use of this formula represents the only genuine difference between the
original slim disc formulation and its time-dependent version considered
here. The original slim disc model used the opacities of Cox and Stewart (1970)
stored in the form of a table. We have checked that stationary models
calculated with these two different opacity treatments do not differ
significantly from each other. 

\subsection {Treatment of the vertical structure}

If the disc is not geometrically thick, it is possible to work in terms of
vertically-integrated variables and to solve an essentially
one-dimensional problem. However, passing from standard to
vertically-integrated variables cannot be done rigorously as the vertical
structure of the disc is not known in detail. Several approximate
procedures have been introduced for dealing with this. We describe here
the approach followed in the present treatment and make comparison with
the alternative approach of Honma, Matsumoto and Kato (1991) (hereafter 
referred to as HMK). 

Neglecting the $zz$-viscous stress tensor component and the vertical 
acceleration term, Eq.~(13) becomes the standard equation of hydrostatic 
equilibrium:

\begin{equation}
{\partial p \over \partial z} = - \rho {\partial \Phi \over \partial z} .
\end{equation}

If one makes the assumption that $p$ and $\rho$ can be linked (in the
vertical direction) by the polytropic relation $p\propto \rho^{1+1/N}$,
where $N$ is the polytropic index, Eq.~(21) can be directly integrated
(Hoshi, 1977) to give

\begin{equation}
\Omega_{_K}^2 H^2 = 2(1+N){p_0 \over \rho_0} ,
\end{equation}
\begin{equation}
\rho (z) = \rho_0\left( {1 - {{z^2}\over {H^2}}} \right) ^N ,
\end{equation}

where the subscript $0$ denotes equatorial plane quantities and
$\Omega_{_K}$ is the Keplerian angular velocity given by $\Omega_{_K} =
\left[ GM/r(r-r_{_G})^2 \right]^{1/2}$. (Hoshi used the Newtonian
gravitational potential rather than the pseudo-Newtonian one but Eqs.~(22)
and (23) are the same in either case.) HMK used $N=3$
for their vertical integrations and in this case, $\Omega_{_K}^2 H^2 =
8p_0 / \rho_0$ but we follow Paczy\'nski and Bisnovatyi-Kogan (1981) and
use

\begin{equation}
\Omega_{_K}^2 H^2 = 6{p_0 \over \rho_0} ,
\end{equation}
as our equation for $H$.

In order to calculate the surface density $\Sigma$, the expression (23)
can be inserted into Eq.~(3) and the integral performed directly. For
$N=3$ one obtains $\Sigma = (32/35)\rho_0 H$ and this was the expression
used by HMK. A very similar result is obtained if $\rho$ is taken to be a
linear function of $z$ with $\rho(0) =\rho_0$ and $\rho(H) =0$. In this
case $\Sigma = \rho_0 H$. If, instead, one took the density to be {\it
constant} in the vertical direction, with $\rho(z)= \rho_0$, then the
result would be $\Sigma = 2\rho_0 H$ which is rather different. The
difference between the first two expressions is small and, given the
approximation introduced when using the polytropic law, there is no
particular reason to prefer one over the other. We use

\begin{equation}
\Sigma = \rho_0 H ,
\end{equation}

for conformity with earlier work. A similar discussion applies for the 
vertically integrated pressure $P$, for which we use

\begin{equation}
P = p_0 H .
\end{equation}

For the radial equation of motion, we use Eq.~(5) with the quantities 
involved taking their equatorial-plane values. HMK, on the other hand, 
used the vertically-integrated form

\begin{equation} 
{Dv_r \over Dt} = 
- {1 \over \Sigma }{\partial P \over \partial r} +
{\left({l^2 - l_{_K}^2}\right) \over r^3}  - {P \over \Sigma}
{d ln \Omega_{_K} \over d r} ,
\end{equation}

which contains the new term $(P / \Sigma)(d ln \Omega_{_K} / d r)$ coming
from the vertical integration of the gravitational potential.  The
question of the importance of this term for time-dependent calculations
has been investigated by Chen (private communication) who concludes that
it does not make any significant difference. 

The thermodynamic relation for the entropy used in our version of the
thermal energy conservation equation is calculated for a mixture of black
body radiation and a simple perfect gas with the polytropic index, $N$,
being set equal to $3/2$. This gives

\begin{equation}
T{DS \over Dt} = {p \over \rho } \left[ (12-10.5\beta) {1 \over T}
{DT \over Dt} - (4-3\beta) {1 \over \rho} {D\rho \over Dt} \right] .
\end{equation}

HMK used the rather similar expression 

\begin{equation}
\rho T {DS \over Dt} = {1 \over \Gamma_3 -1} \left[ {Dp \over Dt} -
\Gamma_1 {p \over \rho} {D\rho \over Dt}\right]
\end{equation} 

which is derived for $c_V =$ constant ($c_V$ being the specific heat at
constant volume). Here $\Gamma_1$ and $\Gamma_3$ are the generalized
adiabatic indices. In their integration procedure, they used the fact that
for $N=3$, $\beta$ does not depend on $z$. The HMK expression for the 
radiative flux $F^-$ (in the integrated energy equation) is 
 
$$ F^- = {16 a c T^4 \over 3 \kappa \rho H}$$

which differs from ours by a numerical factor. 
Comparison between the stationary solutions obtained by Szuszkiewicz
(1988) and those obtained by HMK shows that the results depend only very
weakly (if at all) on the procedure used for the vertical integration
(Abramowicz, Kato and Matsumoto, 1989). It is not necessarily the case,
however, that the same will hold for time-dependent calculations.  Indeed,
Shakura and Sunyaev (1976) pointed out that numerical estimates for
instability growth rates and characteristic wavelengths can be sensitive
to the  procedure adopted for approximating the vertical structure.  

\section{The method used for solving the equations \hbox{numerically}} 

\subsection {The Lagrangian evolutionary code}

The equations presented in Section 2 have been solved numerically, using a
Lagrangian finite difference scheme with standard Lagrangian differencing
and grid organization. The code was adapted from one developed by Miller
and Pantano (1990) in a different context. The integration domain extended
from a suitable position in the supersonic part of the flow (for most of
the models presented here, this was set at $r \approx 2.5\,r_{_G}$) out to
several thousand $r_{_G}$ and was divided into a succession of comoving
annular zones with each one containing a mass 12\% larger than the one
interior to it. The mass of the innermost zone was determined in
accordance with numerical convenience. 
The mass, $m$, interior to a given zone boundary
was used as the radially-comoving independent variable, following the
usual practice for Lagrangian calculations, and the equations of Section 2.1
were rewritten accordingly, with radial gradients being transformed using

$${\partial \over \partial r}=2\pi r \Sigma {\partial \over \partial m}.
$$

During the progress of each calculation, a regridding was carried out
every time that the inner edge of the innermost zone crossed 2.5 $r_{_G}$ in
the course of being accreted. The innermost zone was then removed from the
calculation and all of the variables were interpolated onto a new grid
having a similar   structure (in terms of intervals in the independent
variable $m$) to that at the initial time. The interpolation (which
requires great care in order to avoid introducing destructive
inaccuracies) was carried out using a piecewise cubic method. 
With the grid organization which we use, adequate spatial resolution
is normally obtained with 200 - 300 zones (except in the special cases
mentioned earlier) and the runs have extended for as many as 10$^8$ - 
10$^9$ timesteps. The increase
in mass of the black hole during the calculations, due to accretion of
material from the disc, was extremely small (always less than one part in
$10^{10}$ for the models studied here). 

The equations are solved for $v_r$, $r$, $\Sigma$, $T$ and $l$ as the main
dependent variables. (From this point onwards, the thermodynamic 
parameters $p$, $\rho$ and $T$ will always be taken to refer to values on 
the equatorial plane.) Our numerical scheme has roughly second-order
accuracy in space and in time, the latter being achieved with the use of
intermediate time-levels in a standard way. $\Sigma$ and $T$ are treated
as zone-centre quantities and computed at the full time-level; $r$ is
taken as a zone-boundary quantity, computed at the full time-level and
$v_r$ and $l$ are taken as zone-boundary quantities, computed at the
intermediate time-level. The time-step is adjusted in accordance with the
Courant condition and with two additional constraints on the fractional
variations of $\rho$ and $T$ in any single time-step. 

The continuity equation (2) is rewritten in a form which is more
convenient for the numerical calculations: 

\begin{equation}
{D \over D t}\left( A_{i-1/2} \Sigma_{i-1/2} \right) =0,
\end{equation}

where $i$ is the index of zone boundaries, with $i-1/2$ referring to a 
mid-zone, and $A_{i-1/2}$ is the zone area given by 

$$ A_{i-1/2} = \pi (r_i^2 - r_{i-1}^2 ).
$$

The energy equation (17) is solved implicitly for $T$ by means of an
iterative procedure, using the secant method. Extrapolation of previous
time-step values is used to provide initial estimates, and convergence to
machine accuracy is typically achieved with 4 or 5 iterations. In the
viscous heating term, the gradient of the angular velocity is substituted
by the gradient of the {\it Keplerian} angular velocity, an approximation
which had been used previously in the stationary calculations in order to
avoid a serious numerical instability which otherwise occurred there
(Muchotrzeb and Paczy\'nski, 1982). In the present time-dependent
calculations, we followed the same line in order to remain as close as
possible to the original slim disc approach.

When values are known for $r$, $\Sigma$ and $T$, $p$, $\rho$ and $H$ can
then be found by solving simultaneously the integrated equation of
vertical hydrostatic equilibrium (24), the solution of the integral for
$\Sigma$ (25) and the equation of state (19). These equations are
manipulated algebraically to yield a quadratic equation for $\rho$ which
is then solved analytically.  The solution for $p$, $\rho$ and $H$ is
embedded in the iteration loop for $T$. We take this occasion  to
emphasize that it is necessary to maintain very high levels of accuracy
throughout the various stages of these calculations as there is a very
close balance of terms in both the energy equation and the radial equation
of motion.

A simple form of numerical
diffusion is introduced when solving the radial equation of
motion (5) for $v_r$.  In the finite difference representation of this
equation, $(v_r)_i^n$ ($(v_r)_i$ evaluated at the old time level $t^n$) is
substituted by

\begin{equation}
 \left[ k(v_r)_{i-1}^n + (1-2k)(v_r)_i^n + k(v_r)_{i+1}^n\right]
\end{equation}

with the value $k=10^{-3}$ being selected on the basis of numerical
experiments. The reason for introducing this modification of the standard
scheme will be explained in Section 6.

The inner edge of the grid was placed well within the supersonic part of
the flow, so that the inner boundary conditions could not affect the
evolution external to the sonic point. For sub-Eddington stationary
models, the sonic radius is near to that of the marginally stable orbit at
$r = 3\,r_{_G}$ and so $r = 2.5\,r_{_G}$ is suitable as an initial
location for the inner edge of the grid. During the subsequent evolution,
a check is kept to make sure that the sonic point never gets too close to
the grid boundary. At every time step, it is necessary to set boundary
conditions for $v_r$ and $l$ at both the inner and outer edges of the
grid. At the inner edge, $l$ is set equal to its value at the next grid
point, which is a very good approximation since the specific angular
momentum is very nearly constant in this region, while $v_r$ is calculated
from the standard equation (5) with the pressure gradient set to zero
(which is also an excellent approximation). The outer edge of the grid is
set at several thousand $r_{_G}$ where there is essentially no change in
the variables during the time of the calculation and so $l$ and $v_r$ are
kept constant there.

\subsection{Initial conditions and perturbations}

As initial conditions for the evolutionary calculations, we used the
stationary transonic disc models constructed by Szuszkiewicz (1988) and
described by Abramowicz, Czerny, Lasota and Szuszkiewicz (1988).  The
parameters characterizing these models are: the mass of the black hole
$M$, the accretion rate $\dot M$ which we measure in units of the critical
value $\dot M_c = 64\pi GM/c\kappa$ (the accretion rate for which the 
luminosity is equal to $L_{_E}$), and the viscosity
parameter, $\alpha $. For models with accretion rates up to the critical
one, $\dot M /\dot M_c$ is equal to the luminosity of the disc measured in
units of the Eddington luminosity, $L_{_E} = 4\pi GM m_p c/\sigma_T
\approx 10^{38} (M/M_{\odot})$, erg/s where $m_p$ is the proton mass and
$\sigma_T$ is the Thomson cross section for electron scattering. We will
use $\dot M / \dot M_c$ and $L/L_{_E}$ interchangeably for the
sub-Eddington and Eddington models which are the main focus of our present
investigation.  All of the models discussed here have the same $M$ and
$\alpha$ ($M = 10 M_{\odot}$ and $\alpha = 10^{-3}$) but $\dot M / \dot
M_c$ varies over a wide range of values.

For starting the time-dependent calculations, the initial data from the
appropriate stationary model is first read in from a file and transferred
onto the finite-difference grid. This is a very delicate procedure since,
as mentioned earlier, there is a close balance between large terms both in
the radial equation of motion and in the energy equation. The former is
particularly sensitive. It is important to transfer the minimum possible
number of variables onto the grid and then to calculate other quantities
from the representations of these basic ones on the grid. If this is not
done, inconsistencies arise which destroy the calculation. Despite the
fact that the evolutionary code uses five dependent variables ($r$, $v_r$,
$\Sigma$, $T$ and $l$), it is only necessary to transfer the first four of
these from the stationary model since $l$ can then be calculated from
Eq.~(5) on the grid bearing in mind that, for a stationary model, $Dv_r/Dt
= v_r \, dv_r/dr$. Proceeding in this way turned out to be crucial for the
success of the calculation.

The process of transferring the initial data onto the finite difference
grid together with the action of numerical noise which inevitably enters
during the calculation, turn out to be sufficient for introducing suitable
generalised perturbations into the model to trigger growth of unstable
modes which may be present. A lower limit on the wavelength of
perturbations is, of course, set by the grid spacing but this is amply
fine enough to include the wavelengths of interest.

\section{Global behaviour of the models}

We have found three different types of global behaviour for the present
initially stationary slim disc models depending on the accretion rate or,
equivalently, on the luminosity. Models with $L \le 0.08\,L_{_E}$ are
stable. Models with $L$ between 0.09 and 1$\,L_{_E}$ develop a violently
unstable shock-like feature near to the sonic point, which is thought to
lead to disruption of the disc, and our calculation then terminates.
Finally, models with super-Eddington luminosities show progressive slow
evolution. We concentrate here on the sub-Eddington models, leaving
discussion of the super-Eddington ones for a forthcoming paper. 

As background for our discussion of the results of the time-dependent
calculations, it is useful to review briefly the predictions of local
stability analysis carried out for stationary models.  The simplest local
stability criterion states that if, at a certain radius, the ratio of gas
pressure to total pressure, $\beta$, is smaller than 0.4, then the disc
will be unstable at that radius. We have constructed a sequence of
stationary models, with a range of luminosities, and key parameters are
listed in Table 1. The local stability criterion has been applied at every
radius for each model and the locations of predicted unstable regions have
been determined. If, instead, we use the refined criterion derived by
Pringle (1976), the predicted unstable regions remain essentially the same
for all of the models except the one with $L = 0.07 \, L_{_E}$ for which
the range becomes $6.2 - 9.0 \, r_{_G}$. This small difference is due to
Pringle's result that when electron scattering does not provide the
dominant opacity, the disc is thermally stable even when the dominant
contribution to the pressure comes from radiation. The models with
luminosity less than $0.07\,L_{_E}$ are predicted to be stable on the
basis of local analysis. For $L \ge 0.07 \, L_{_E}$, the disc can be
divided into three regions: an inner locally stable zone extending from
the sonic point (at $r_{sonic}$) out to the inner edge of the unstable
region, the unstable zone (the location of which is given in Table 1) and
finally an outer stable zone. 

Following Shakura and Sunyaev (1976), it is possible to give a more
extended description of the stability properties of our stationary models,
still on the basis of local linear analysis. While this description is
somewhat incomplete and has limitations, it is nevertheless quite
instructive. For each model, we evaluate the bifurcation wavelength, which
we will call $\Lambda _{min}$. Perturbations with wavelengths, $\Lambda $,
satisfying $2H < \Lambda < \Lambda _{min} $ give concentric waves
travelling across the surface of the disc while for $\Lambda \ge \Lambda
_{min}$ there are two branches of growing modes which take the form of
standing waves. For the first of these branches, the perturbation growth
rate decreases with increasing wavelength and in the limit $\Lambda \gg H$
this becomes the instability discovered by Lightman and Eardley (1974). On
the second branch, the growth rate {\it increases} with increasing
wavelength and growth of these modes is due to the thermal instability
found by Pringle, Rees and Pacholczyk (1973). The characteristic growth
time, given by Shakura and Sunyaev, is

\begin{equation}
\tau= {56-45\beta -3\beta ^2 \over 30(0.4-\beta)}t_{th} ,
\end{equation}

with the thermal time scale $t_{th}$ being given by  

$$ 
t_{th} = {1 \over \alpha \Omega} .
$$

The growth time, $\tau$, is different for different radii. In Table 1 we
give its minimum value for each model, $\tau_{min}$, and the location in
the disc where this minimum is attained, $r(\tau_{min})$. Each of the
stationary models listed was evolved forward in time with our
time-dependent code, as described earlier, and the final column of the
table shows the (physical) time for which each of these calculations was
continued. This represents either the time at which the calculation was
discontinued because of catastrophic instability growth or the time beyond
which it was no longer considered to be of interest to continue.

The aim of our time-dependent calculations was to determine which models
are {\it globally} unstable and to see how the instabilities evolve with
time. Firstly, as a test of the code, we calculated the evolution of the
model with $L = 0.01 \, L_{_E}$ which is {\it stable} according to the
linear analysis. Without modification, the original basic code showed even
this model becoming unstable but we found that the instability could be
removed by introducing a very small amount of additional numerical
diffusion into the scheme, following the prescription of Eq.~(31). The
additional diffusion was then retained for all of the models studied. We
will discuss this further in \hbox{Section 6.}

Models with $L=0.07$ and $0.08\,L_{_E}$ were not found to develop any
global instability, a result which could already be explained within the
context of the local analysis since the minimum wavelength
$\Lambda_{min}$, necessary for the onset of the thermal instability, is
larger than the width of the unstable region for these models.

For $L$ in the range from 0.09 to $1\,L_{_E}$, the evolution is terminated
by the formation of a narrow velocity spike adjacent to the sonic point
which, once initiated, grows very rapidly in amplitude and disrupts the
solution. As a representative example of these cases, the global behaviour
of the model with $L=0.1\,L_{_E}$ is illustrated in Figures 1, 2, 3 and 4
which show the radial profiles of the radial velocity divided by the sound
speed (i.e. the Mach number), the surface density, the temperature and the
thickness of the disc, at several time-levels during the evolution. An
initial instability begins to grow gradually in the region predicted to be
unstable by the local linear analysis and the influence of this then
spreads inwards to the sonic point, where it triggers a second, more
violent, instability (as shown in Figure 5). This is a fundamentally
non-local process. For this particular model, the perturbation in
$v_r/c_s$ associated with the initial instability extends outwards to
beyond $10\,r_{_G}$ and, by the time that the run is ended, it has
saturated in amplitude and a small region of positive (outward) radial
velocity has appeared, peaking at $\sim 7\,r_{_G}$. There are
corresponding variations in the other parameters, which are in the sense
of increasing temperature and disc thickness and decreasing surface
density in the region of the initial instability. 
The sonic point moves progressively inwards (despite what might be inferred
from Figure 1). As will be seen from the
local views of the global behaviour presented in Section 5, the local
accretion rate at around $3\,r_{_G}$ increases more quickly than that at
$5\,r_{_G}$ and this leads to an emptying of the inner part of the disc.
For $L=0.1\,L_{_E}$, the velocity spike near to the sonic point starts to
appear almost simultaneously with the saturation in amplitude of the
initial perturbations in $v_r/c_s$ and $T$ and, once initiated, the growth
occurs very rapidly (in $\sim 7 \times 10^{-3}$ s, i.e. on the dynamical
timescale) with a shock-like
structure developing at the leading edge of the spike. There are
corresponding rapid changes in the profiles of the other variables. The
growth of this feature appears to be catastrophic and, certainly, we are
not able to continue the evolution further with our present numerical
scheme. However, we are not able to completely rule out the possibility
that there might be a stabilization in the non-linear regime which could
be followed with a more sophisticated numerical treatment.  We have checked, a
posteriori, that even under these most extreme conditions the models
remain optically thick, according to the criterion $(\kappa_e \kappa
_{ff})^{1/2} (\Sigma/2)>1$, both in the expanded innermost part and in the
region of the velocity spike. Since we strongly suspect that the velocity
spike will be suppressed or severely altered as a result of including
improvements to  the original slim disc model, we delay further 
consideration of its origin and behaviour until a subsequent paper
where the effects of such improvements will be systematically discussed.

While the case $L=0.1\,L_{_E}$, discussed in detail above, is largely
representative of the others with $L$ in the range from 0.09 to
$1\,L_{_E}$, there is a difference appearing for models with slightly
higher luminosity which needs to be mentioned here.  With the case just
considered, the run is halted by the growth of the velocity spike almost
immediately the saturation occurs in the initial perturbation. However, it
can be seen (particularly in Fig.~2 for $T$) that the feature associated
with the initial instability is then just starting to propagate outwards.
In higher luminosity models, such as the one with $L=0.2\,L_{_E}$ which we
will discuss in the next section, this feature has time to propagate
outwards significantly before the run is ended. It is tempting to
speculate that this might, perhaps, have been the beginning of a cyclic
behaviour if the instability near to the sonic point had not intervened.

\section{Local view of the global behaviour - S-curves and evolutionary
tracks } 

The existence of S-curves in the $\log \dot M - \log \Sigma$ plane for
sequences of models of stationary accretion discs around black holes,
provided a motivation to look for nonlinear cyclical behaviour in these
systems in analogy with the situation for dwarf novae. The local stability
of a disc at a given radius, $r$, is related to the location of the model
on the S-curve for that radius. The lower and middle branches of the
S-curve correspond to the standard Shakura/Sunyaev models and are simply
loci representing the balance between viscous heating and radiative
cooling. Discs on the lower branch are thermally stable while those on the
middle branch are unstable and a transition via marginal stability occurs
at the lower turning point. The upper branch has a positive slope which
might indicate stability, but here the situation is different from that
for dwarf novae. On the upper branch here, the heat generated by viscosity
is mainly advected with the accretion flow, rather then being mainly
radiated away as on the lower branches, and standard local analysis is not
adequate for determining whether these models are stable or not. 

Having made our global calculations as described in the previous section,
we can then take a local view of them, considering the time variation of
parameters at a particular fixed location $r$. Following earlier work on
local analysis of stationary models, it is convenient to consider plots of
$\log \dot m$ against $\log \Sigma$ but it should be noted that the
trajectories exit from these plots whenever there is a local outflow,
$\dot m < 0$. For the case shown in Fig.~1 ($L = 0.1\, L_{_E}$), the local
outflow is seen only at the very end of the run but for models with higher
luminosity, the outflow feature can sometimes be seen passing by the
monitoring location. In Figure 6, we show the trajectory of parameters at
$r = 8\,r_{_G}$ for a model with $L = 0.2\,L_{_E}$. (Here and in the rest
of this section, $\dot m$ is measured in units of the critical accretion
rate, $\dot M_c$.) The dashed lines mark rapid motion out of (or into) the
parameter space, corresponding to $\dot m$ changing sign. In this case,
$\dot m$ remained negative for 14 seconds before becoming positive again.

This figure is the first of a sequence of similar ones showing different
models and monitoring locations and so we take this opportunity to
describe the way in which all of them should be viewed. Stationary models
covering the whole range of luminosities are located along the S-shaped
curve shown by the dotted line and we plot the phase-space trajectory for
the particular model of interest at the selected monitoring location
together with corresponding trajectories for a set of comparison models
having initial $\dot m = 0.01$, 0.07, 0.08, 1 and 10. We recall that for
the initial stationary models, $\dot m(r) = \dot M = const$ and $\dot m$
measured in units of $\dot M_c$ is equal to $L/L_{_E}$ for those models
which are not super-Eddington. Identification of the different initial
models is straightforward as it is sufficient to look at the corresponding
value of $\dot m$ on the axis. For each model, the open square shows the
initial state and the filled square shows the state at the end of the
calculation with the line joining them indicating the phase space
trajectory followed. Models for which it is not possible to see any change
in location on the scale of this plot have the filled square superimposed
on the open one so that only the filled square is seen.  Models with
initial $\dot m = 0.01$, 0.07 and 0.08 are globally stable and always
remain at essentially the same point in the phase space for the given
radius. The model with Eddington luminosity, $\dot m = 1$, evolved for
only a very short time before the run terminated and so it is only for
small radii (where the evolutionary time-scale is shorter) that any
movement can be seen in the phase space. Finally the super-Eddington
model, which we are not discussing in detail here, shows similar sorts of
trajectory at the different radii. 

Although local analysis for discs around neutron stars and black holes is
normally discussed in terms of the $\dot m - \Sigma$ relation, most
authors presenting results from time-dependent calculations have shown
local views in the $\log T - \log \Sigma$ plane instead of the $\log \dot
m - \log \Sigma$ one. This is reasonable since $T$ and $\Sigma$ are
primary quantities calculated in the numerical codes, whereas $\dot m$ is
a secondary quantity used just for interpretation, and also there is the
problem mentioned earlier, that negative values of $\dot m$ cannot be
represented in a plot of $\log \dot m$ against $\log \Sigma$. In Figure 7
(which focuses on our ``standard'' model with $\dot m = 0.1$) we show 
trajectories in both planes for comparison. Four different monitoring
locations are used (at 3, 5, 8 and 10$\,r_{_G}$) and the frames are
arranged in two columns, one for $\log \dot m - \log \Sigma$ and the other
for $\log T - \log \Sigma$. The general conclusions which can be drawn from
considering the qualitative differences between the evolutionary tracks
for different radii, are that different parts of the disc do not react in a
synchronized way and that the instability behaviour seen has a global
character. Our discussion will mainly be related to the $\log \dot m - 
\log \Sigma$ curves. 

For $r = 3\,r_{_G}$, viewed in terms of $\log \dot m - \log \Sigma$, the
model evolves along the lower branch of the S-curve for $178\,$s but then
overshoots the position where the S-curve bends and makes an attempt to
come back again. However, soon after it turns and changes its direction,
it starts to follow a quite dramatic path and the run is halted soon after
this. We know, from the global behaviour, that this is caused by the
development of the instability at the sonic point. The evolutionary track
in the $\log T - \log \Sigma$ plane is less dramatic, but it is clear also
from this that the evolution is not going to follow any closed path as
would be necessary for having a limit cycle behaviour. The situation at
5$\,r_{_G}$, if viewed in isolation, would have seemed promising for
producing a limit cycle behaviour, especially looking at the $\log T -
\log \Sigma$ plane where the model follows a well-behaved regular path,
but it is important to notice that the local accretion rate here grows
significantly less than that at $3\,r_{_G}$ giving rise to an unbalanced
situation producing a drop in surface density.  The local accretion rates
at 8 and $10\,r_{_G}$ decrease in time, making the effect even more
pronounced. This description complements the global view of the behaviour
of the surface density shown in Figure 3. 

For models with higher accretion rates, the importance of the transonic
nature of the flow, deviation away from Keplerian rotation and advective
cooling becomes progressively greater. To investigate how these non-local
effects influence the evolution of models with higher accretion rates, we
made calculations for several models with luminosities between 0.1 and
$1\,L_{_E}$. In Figure 8, we show the evolutionary paths for $L = 0.2$,
0.4 and $0.8\,L_{_E}$ at 3 and $5\,r_{_G}$. The evolution for $L=
0.2\,L_{_E} $ (which we have already discussed in connection with Fig.~6,
drawn for $8\,r_{_G}$) proceeds in a rather similar way to that for
$0.1\,L_{_E}$ except that the velocity perturbation is here able to
propagate outwards significantly and the run terminates with unstable
behaviour at the sonic point after a shorter time ($71\,$s in this case).
These features might seem contradictory at first sight but it should be
noted that the radial velocities are higher for $0.2\,L_{_E}$ and so
everything proceeds more rapidly. Moving to models with higher
luminosities, already for $L=0.4\,L_{_E}$ the situation becomes
significantly different. Now, the local accretion rate increases
systematically at {\it all} of the radii considered in Fig.~7. The Mach
number perturbation propagates out to around $25\,r_{_G}$ but never gives
any local outflow. For $L=0.8\,L_{_E}$ the Mach number perturbation does
not have enough time to propagate outwards before the unstable behaviour
at the sonic point terminates the run (after $25\,$s). This is directly
connected with the fact that the drift time $\,t_{dr}=r/v_r$ (which
becomes progressively shorter for models with higher accretion rates) is
now comparable with the thermal time-scale (Szuszkiewicz, 1990). By this
stage, it is no longer the case that the thermal instability determines
the evolution of the disc. 

Our results indicate that the models presented here with luminosities in
the range 0.09 - $1\,L_{_E}$ are globally unstable and that the
demarcation line between globally stable and globally unstable models lies
between 0.08 and $0.09\,L_{_E}$. It is interesting to compare the
trajectories in the $\log T- \log \Sigma $ plane for models on either side
of the demarcation line. Figure 9(a) shows trajectories for two models above
it:
0.09 and $0.1\,L_{_E}$, at $r = 8\,r_{_G}$. We focus attention on the one
with $0.09\,L_{_E}$ (which is nearer to the demarcation line) and note the
evolution starting along an unwinding spiral and then diverging off
towards higher temperatures. From the corresponding time series for the
temperature variations, shown in Figure 9(b), we see the temperature
starting to oscillate with growing amplitude but then quickly turning to
unbounded growth.  Figure 10(a) shows the trajectory in the $\log T - \log
\Sigma $ plane for the model with $L=0.08\,L_{_E}$ at $r = 8\,r_{_G}$. The
evolution proceeds along the inspiralling path from the initial, slightly
perturbed, state to the asymptotic one with only very small changes in the
temperature and surface density. (All of this, and the associated
behaviour in the $\log \dot m - \log \Sigma $ plane, was hidden inside the
filled squares in Figures 6, 7 and 8.) The corresponding time series is
shown in Figure 10(b) where one can see oscillations being damped 
to zero amplitude. 

\section {Comments and conclusions }

In this section, we would first like to comment on issues connected with
our introduction of additional  numerical diffusion. This was motivated by
observing that, in our first calculations, the supposedly stable model
with $L=0.01\,L_{_E}$ did not remain unchanging but instead
developed an instability in the region $3 - 6\,r_{_G}$ after $10-20\,$s of
the evolution. This appeared first at around $4.7 - 4.8\,r_{_G}$ and 
grew rapidly, causing termination of the calculation. Figure 11 shows the
velocity and surface-density profiles for the developed instability. This
behaviour was not entirely unexpected since it was already known that
additional global instabilities may be introduced as a direct consequence
of literal adoption of the Shakura/Sunyaev $\alpha$ viscosity 
prescription (see, for example, Abramowicz, Papaloizou and Szuszkiewicz,
1993). However, it was important to be sure that the behaviour seen was
not merely an artefact of our numerical scheme and so we carried out
several experiments to check on this including using an implicit method
for the radial velocity equation and a three-point formula for the
pressure gradient (the regridding procedure had already been subjected to
extensive testing). The instability persisted unchanged in each case.
Another set of experiments was carried out to test the effect of varying
the viscosity parameter $\alpha$. The instability was always present but
appeared at different locations and after different times. We also tried
different viscosity formulations but found that the instability was always
present. F. Honma (private communication) has performed a calculation for
the same model with his code but did not find the same behaviour. Our
suspicion is that the discrepancy arises because of the fact that our
basic code is very non-diffusive and, bearing in mind that the standard
slim-disc model does not include  several potentially important diffusive terms,
we decided to try adding a low level of numerical diffusion to the radial
equation of motion, as discussed earlier. We found that the instability
was suppressed even with extremely small values of $k$, suggesting that
the additional diffusive terms discussed in Section 2, but not included in
the present formulation, probably play a crucial role for avoiding this
instability which would otherwise make these models unsuitable as
representations of real astrophysical discs. Since we wanted to focus on
the more interesting non-stationary behaviour described in the main part
of this paper, we included the additional diffusion for all of our
calculations to avoid them being terminated prematurely. Without doing
this, all of our sub-Eddington models suffer in the same way as the one
with $L=0.01\,L_{_E}$ but no similar behaviour is seen for super-Eddington
models. We emphasize
that the kind of numerical scheme which we are using here has been
extensively tested for a wide range of physical situations and does not
normally require the addition of extra numerical diffusion to give stable
behaviour in circumstances such as these. 

The aim of this paper has been to study the nonlinear evolution of the
thermal instability for the original version of the slim disc model 
 and to verify predictions concerning the
stabilizing effect of advection.  Setting aside the instability discussed
in the previous paragraph, we have found that models predicted to be
stable by local analysis do remain stable and stationary. The same 
applies for models which, according to the local analysis, would have a
potentially unstable region smaller than the minimum wavelength for
unstable perturbations. In terms of luminosities, this means that all
models with luminosity less than or equal to $0.08\,L_{_E}$ are globally
stable. For models with luminosities between $0.09\,L_{_E}$ and
$1\,L_{_E}$, we see the run ended by  unstable behaviour near to the
sonic point. 

For the models studied here and the particular formulation adopted, it
seems very unlikely that advection alone can be sufficient to allow for
the existence of limit cycle behaviour. 
Non-local effects which are not included here (such as radiative
diffusion in the radial direction) might well play 
important role in the global
evolution of the thermal instability since they introduce parabolic terms
which could significantly affect the delicate radial force-balance.

Our results are rather different
 from those reported  by the Japanese group (HMK).  
However, it should be emphasized that
they  used a viscosity parameter two orders of magnitude larger than
ours and this gives rise to models which are very different from the
original slim disc formulation which we have been studying here. In
particular, it means that they are dealing with much larger  radial
velocities and hence the effect of advection is much stronger. 
We are currently in the process of investigating models with larger
$\alpha$. 

In our future work, we plan to follow the programme outlined at the
beginning of the paper: namely, to use the present study as a standard
against which to judge the effects of systematically adding further
sophistications onto the original slim disc formulation. Also, we have
been studying super-Eddington models. A global analysis is really
essential for discussing the stability of these since the local
approximation is rather poor in this case. However, some qualitative
predictions have been made on the basis of local analysis (Abramowicz et
al. 1988, Wallinder 1991) and it will be interesting to compare them with
the calculated global behaviour. 

\section* {Acknowledgments}

We would like to thank Marek Abramowicz, Bo\.zena Czerny, Fumio Honma,
Shoji Kato, Ryoji Matsumoto and  John Papaloizou for some very interesting 
discussions  about our results. 

We gratefully acknowledge financial support from the U.K. Particle Physics
and Astronomy Research Council and the Italian Ministero dell'Universit\`a
e della Ricerca Scientifica e Tecnologica. 

\newpage
\parindent 0pt
\parskip 11pt

\section* {References}
 
\hangindent 1.0 truecm
Abramowicz, M.A., Czerny, B., Lasota, J-P., Szuszkiewicz, E., 1988,
ApJ, 332, 646 

Abramowicz, M.A., Kato, S., Matsumoto, R., 1989, PASJ, 41, 1215

\hangindent 1.0 truecm
Abramowicz, M.A., Papaloizou, J.C.B., Szuszkiewicz, E., 1993,
Geophys. Astrophys. Fluid Dyn., 70, 215

Chen, X., Taam, R.E., 1993, ApJ, 412, 254

Chen, X., Taam, R.E., 1994, ApJ, 431, 732 

Cox, A.N., Stewart, J.N., 1970, ApJ Suppl., 19, 243 

Honma, F., Matsumoto, R., Kato, S., 1991, PASJ, 43, 147 

Honma, F., Matsumoto, R., Kato, S., Abramowicz, M.A., 1991, PASJ, 43, 
147 

Hoshi, R., 1977,  Prog. Theor. Phys., 58, 1191

Kato, S., Abramowicz, M.A., Chen, X., 1996, PASJ, 48, 67

Lasota, J.-P., Pelat, D., 1991, A\&A, 249, 574

Lightman, A.P., 1974, ApJ, 194, 429 

Lightman, A.P., Eardley, D.M., 1974, ApJ, 187, L1
 
Loska, Z., 1982, Acta Astron., 32, 13

\hangindent 1.0 truecm
Matsumoto, R., Kato, S., Honma, F., 1989, in Meyer, F., Duschl, W.J.,
Frank, J., Meyer-Hofmeister, E., eds, Theory of Accretion Disks. Kluwer
Academic Publications, London, p. 167

Meyer, F., and Meyer-Hofmeister, E., 1982, A\&A, 106, 34

Miller, J.C., Pantano, O., 1990, Phys. Rev. D, 42, 3334 

Muchotrzeb, B., Paczy\'nski, B., 1982, Acta Astron., 32, 1

Paczy\'nski, B., 1980, Acta Astron., 30, 347 

Paczy\'nski, B., Bisnovatyi-Kogan G., 1980, Acta Astron., 31, 283

Paczy\'nski, B., Wiita, P.J., 1980, A\&A, 88, 23

Papaloizou, J.C.B., Faulkner, J., Lin, D.N.C., 1983, MNRAS, 205, 487

Papaloizou, J.C.B., Stanley, G.Q.G., 1986, MNRAS, 220, 593

Papaloizou, J.C.B., Szuszkiewicz, E., 1994, MNRAS, 268, 29

Piran, T., 1978, ApJ, 221, 652 

Pringle, J.E., 1976, MNRAS, 177,65 

Pringle, J.E., 1981, Ann. Rev. Astron. Astrophys., 19, 137

Pringle, J.E., Rees, M.J., Pacholczyk, A.G., 1973, A\&A, 29, 179

Shakura, N.I., Sunyaev, R.A., 1973, A\&A, 24, 337

Shakura, N.I., Sunyaev, R.A., 1976, MNRAS, 175, 613

Smak, J., 1982, Acta Astron., 32, 199

Szuszkiewicz, E., 1988, Ph.D. Thesis, SISSA, Trieste  

Szuszkiewicz, E., 1990, MNRAS, 244, 377 

Taam, R.E., Lin, D.N.C., 1984, ApJ, 287, 761

Wallinder, F.H., 1991, A\&A, 249, 107  

\newpage
\parindent 12pt
\parskip 20pt
\noindent 
{\bf Table 1:} Results from local stability analysis of the initial models \\
and the total physical time for which each model was evolved.

\noindent 
\begin{tabular}{|c|c|c|c|c|c|c|}     \hline
  &  &  &  &  &  &\\
${L / L_{_E}}$ & 
$r_{sonic}$ &
unstable region   & 
$\Lambda  _{min}$ &
$ r(\tau _{min})$  & 
$\tau_{min}$ & $t_{calc}$ \\  
  &  &  &  &  &  &\\
 & [$r_G$]& [$r_G$] & $[r_G]$ & [$r_G$] &[s]&[s] \\
  &  &  &  &  &  &\\  \hline
  &  &  &  &  &  &\\
0.01 &2.96 & stable everywhere   &  & & & \\ 
  &  &  &  & & & \\
0.07 & 2.95 & 6.2 - 9.7  & 28.6  & 7.4 & 226 & $>$ 5650 \\ 
  &  &  &  & & & \\
0.08 & 2.95 & 5.3 - 12.8  & 7.7  &6.6 & 56.8 & $ >$ 7120 \\ 
  &  &  & &  & &  \\
0.09 & 2.95 & 4.9 - 15.2 & 4.7  &6.3 & 33.4 & 582 \\ 
  &  &  & &  & &  \\
0.1 & 2.94  & 4.6 - 17.5 & 3.6  &5.9 & 24.3 & 182 \\ 
  &  &  & &  & &  \\
0.2 & 2.94  & 3.8 - 37.1 & 1.6  &4.8 & 8.5 & 71 \\ 
  &  &  & &  & &  \\
0.4 & 2.93 & 3.4 - 70.7 & 1.1 &4.1 &5.1 & 126 \\ 
  &  &  & &  & &  \\
0.5 & 2.93  & 3.3 - 86.1 & 1.0  &3.9 & 4.5 & 155 \\ 
  &  &  & &  & &  \\
0.6 & 2.92  & 3.3 - 100.8 & 0.9  &3.7 & 4.0 & 165 \\ 
  &  &  & &  & &  \\
0.7 & 2.92  & 3.2 - 115.1 & 0.8  &3.5 & 3.6 & 135 \\ 
  &  &  & &  & &  \\
0.8 & 2.91 & 3.1 - 129.0 & 0.7 &3.4 &3.3 & 25\\ 
  &  &  & &  & &  \\
1.0 & 2.82 & 2.9 - 155.8  & 0.5 &3.1 &2.7 & 0.08 \\ 
  &  &  & &  & &  \\
  &  &  & & & &  \\ \hline
\end{tabular}

\newpage
\parindent 0pt
\parskip 11pt
 
{\bf Figure Captions}

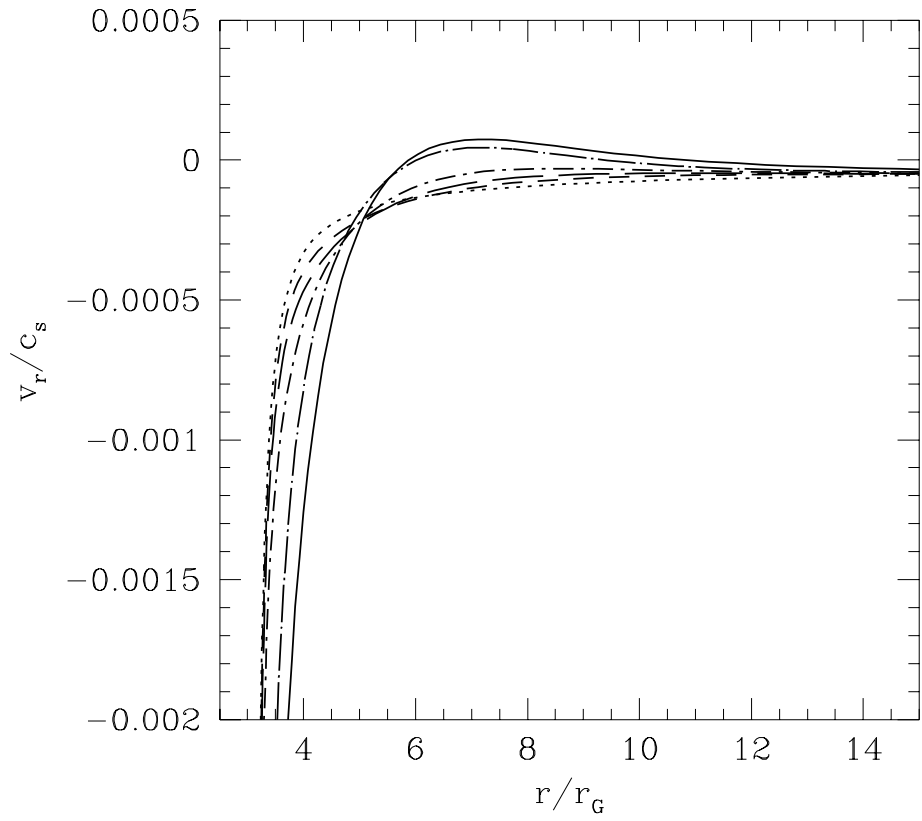
{\bf Figure 1:} Mach number ($v_r/c_s$) plotted as a function of
$r/r_{_G}$ for the model with $L=0.1\,L_{_E}$ at several different times:
$t=0\,$s (dotted line), $t=150\,$s (short-dashed line), $t=162\,$s 
(long-dashed line), $t=171\,$s (dot/short-dashed line), $t=178\,$s
(dot/long-dashed line), and $t=180\,$s (solid line). 

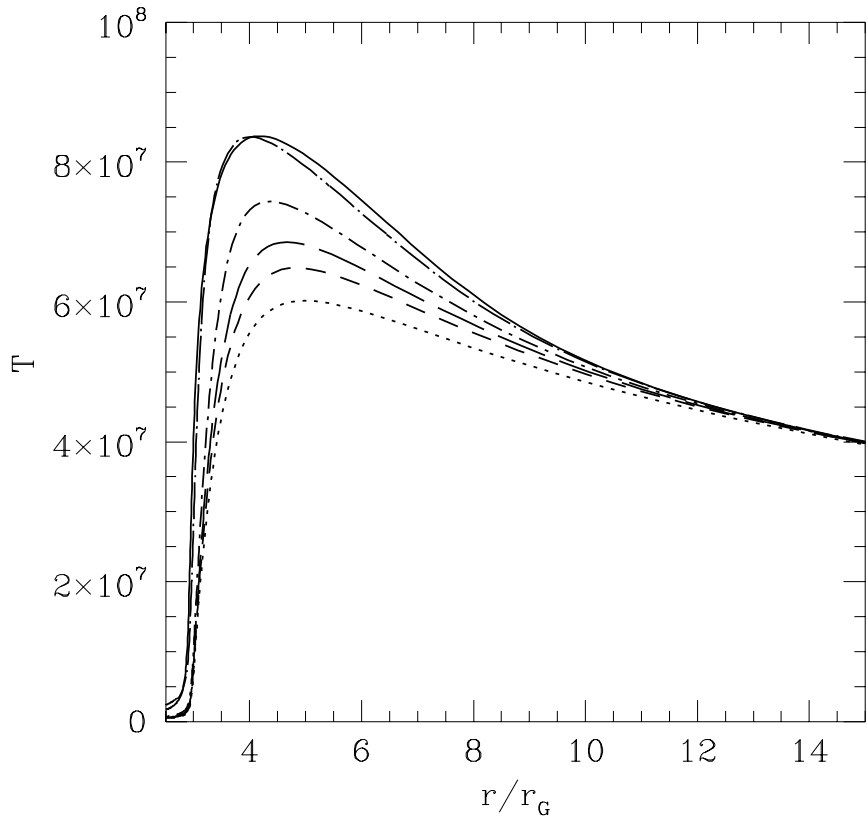
{\bf Figure 2:} Temperature $T$ (in degrees Kelvin) plotted as a
function of $r/r_{_G} $ for the model with $L=0.1\,L_{_E}$ at several 
different times. The line description is the same as for Figure 1. 

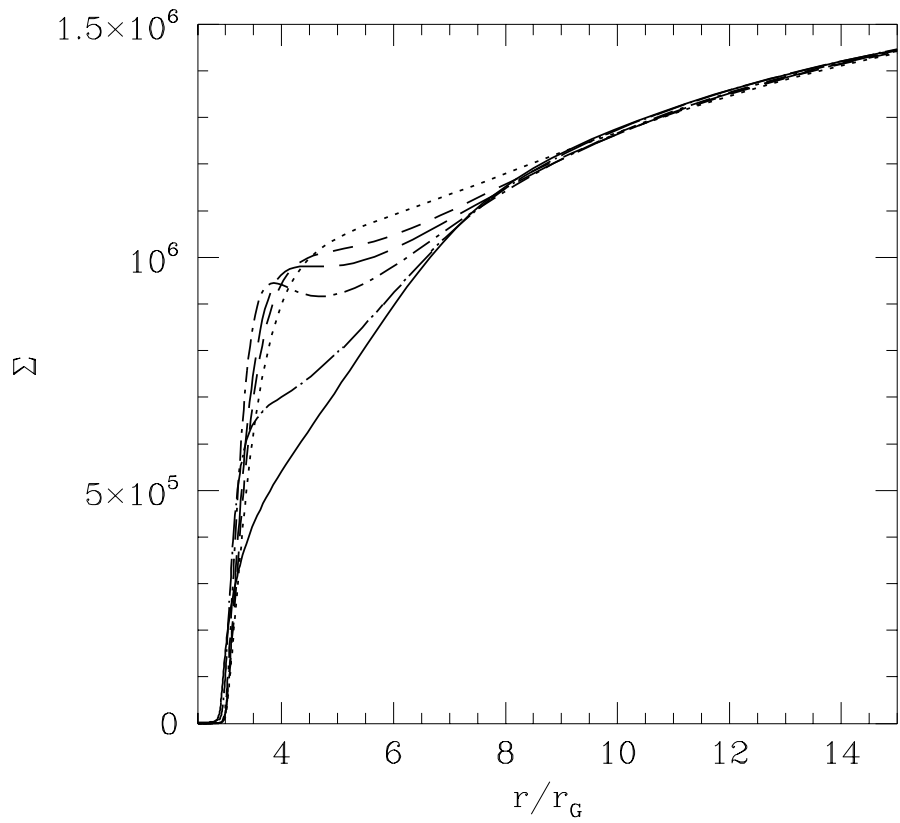
{\bf Figure 3:} Surface density $\Sigma$ (in g/cm$^2$) plotted as a
function of $r/r_{_G} $ for the model with $L=0.1\,L_{_E}$ at several 
different times. The line description is the same as for Figure 1.

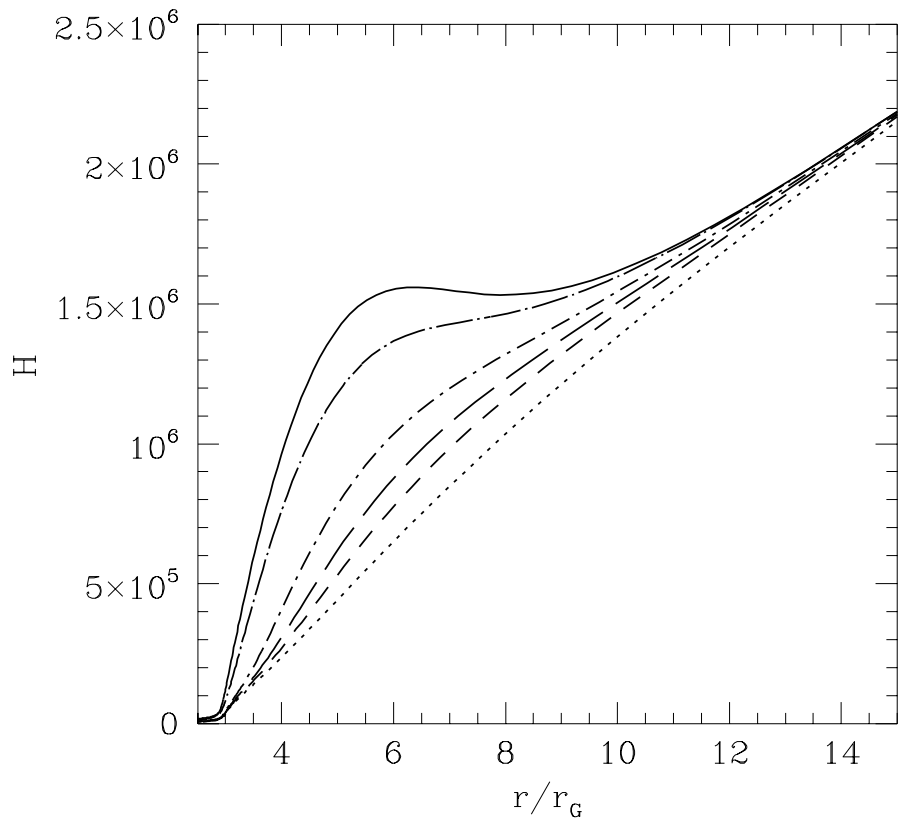
{\bf Figure 4:} The half-thickness of the disc, $H$ (in cm) plotted as a
function of $r/r_{_G} $ for the model with $L=0.1\,L_{_E}$ at several 
different times. The line description is the same as for Figure 1.

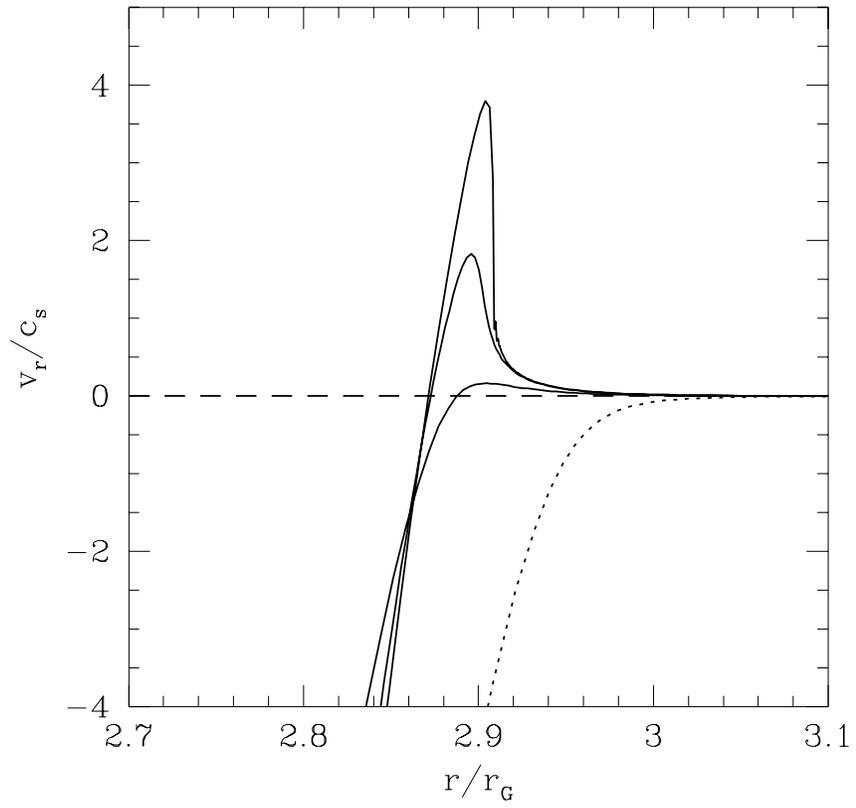
{\bf Figure 5:} Development of the instability near to the sonic point 
($v_r/c_s= -1$) shown in terms of the Mach number. The model concerned is
again the one with $L=0.1\,L_{_E}$. The dotted line shows the initial Mach
number profile and the rapid growth of the sharp-peaked structure is shown
by profiles for three consecutive output times: $t=180.2001\,$s, 
180.2028$\,$s and 180.2069$\,$s. 

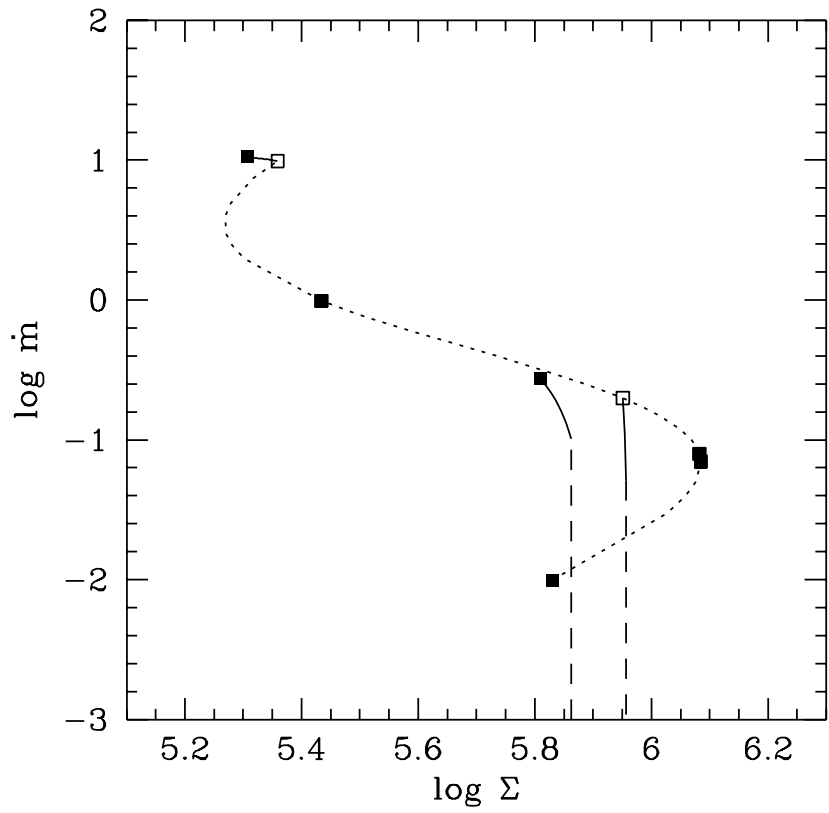
{\bf Figure 6:} The $\log \dot m - \log \Sigma$ plane for $r = 8\,r_{_G}$
showing the location of the sequence of stationary models (dotted line)
and the evolutionary paths (solid and dashed lines) for models with $\dot
m = 0.01$, 0.07, 0.08, 0.2, 1 and 10. The initial location for each model
is marked by an open square and the final position is marked by a filled
square. See text for a more detailed description. 

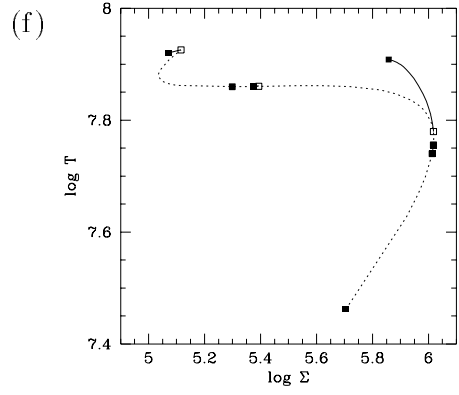
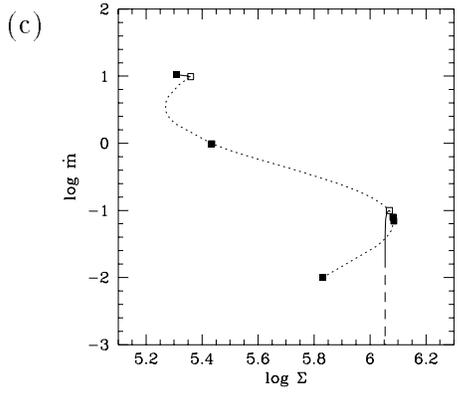
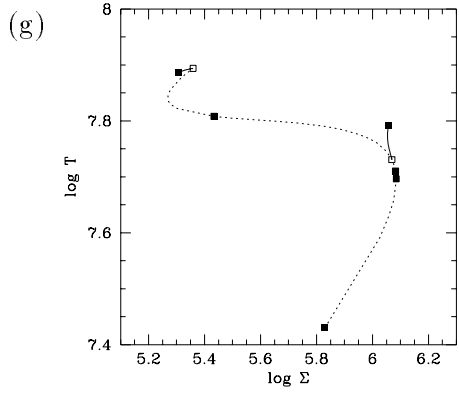
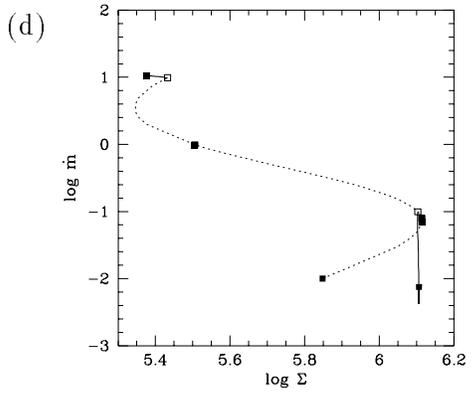
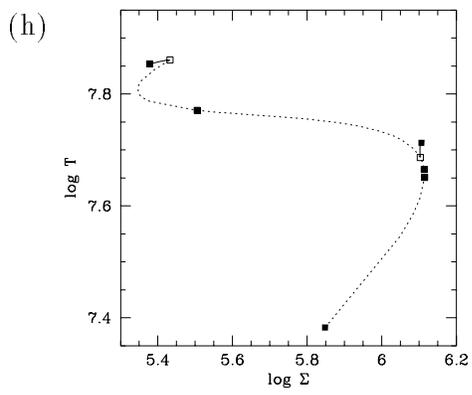
{\bf Figure 7:} The left-hand column shows the $\log \dot m - \log \Sigma$
plane for four different radii: (a) 3$\,r_{_G}$, (b) 5$\,r_{_G}$, (c)
8$\,r_{_G}$ and (d) 10$\,r_{_G}$. The evolutionary paths are for the same
models as in Fig.~6 except that $\dot m = 0.2$ is replaced by $\dot m =
0.1$. The right-hand column (figures (e) -- (h)) shows the corresponding
behaviour in the $\log T - \log \Sigma$ plane. 

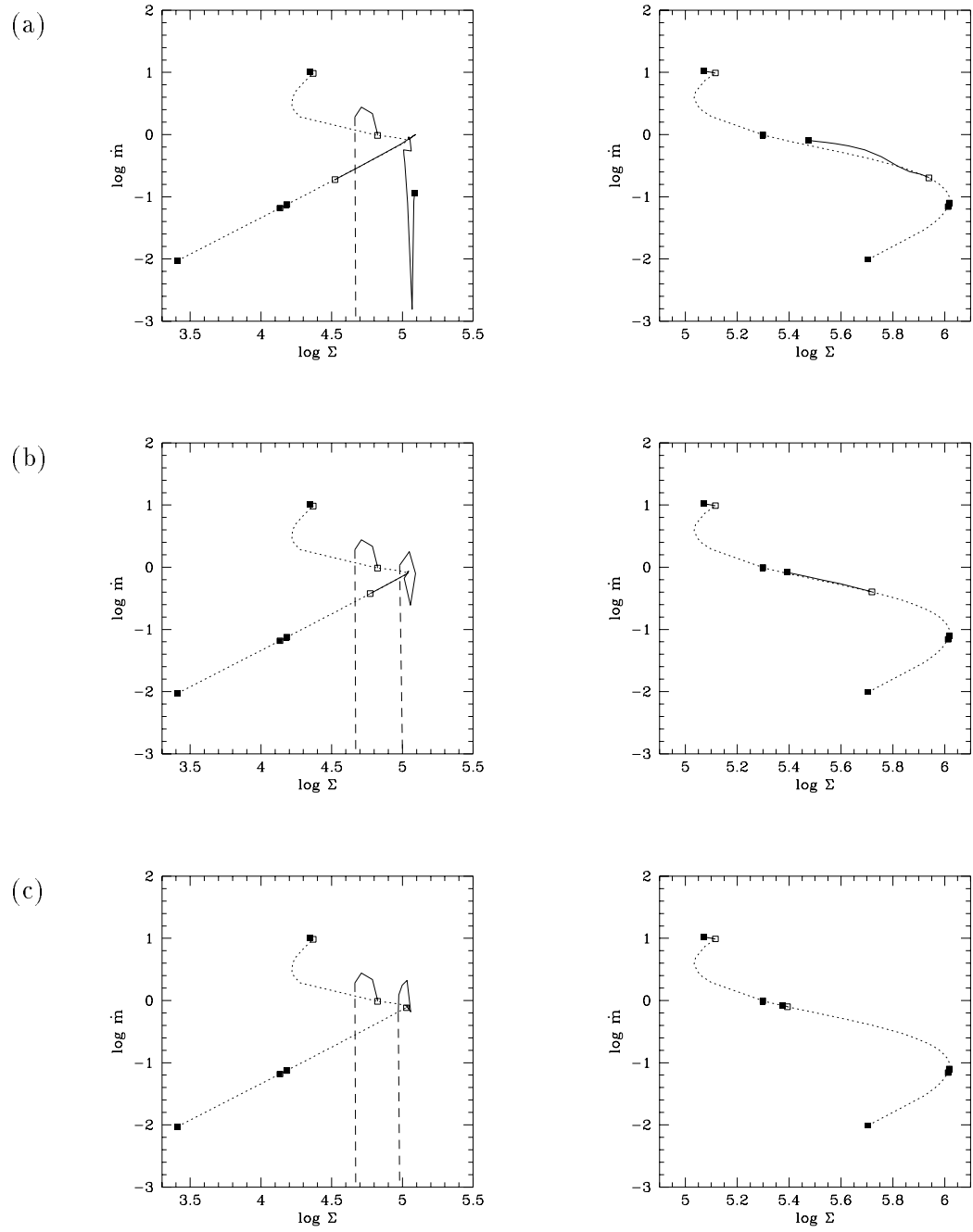
{\bf Figure 8:} The $\log \dot m - \log \Sigma$ plane for 3$\,r_{_G}$ 
(left-hand column) and 5$\,r_{_G}$ (right-hand column) for three different 
models: (a) $\dot m = 0.2$, (b) $\dot m = 0.4$ and (c) $\dot m = 0.8$.

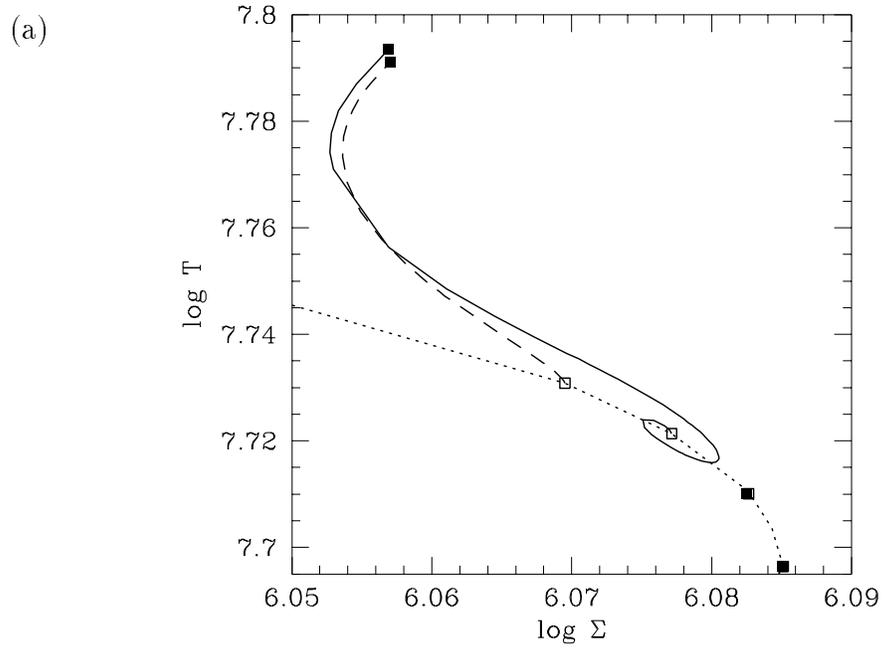
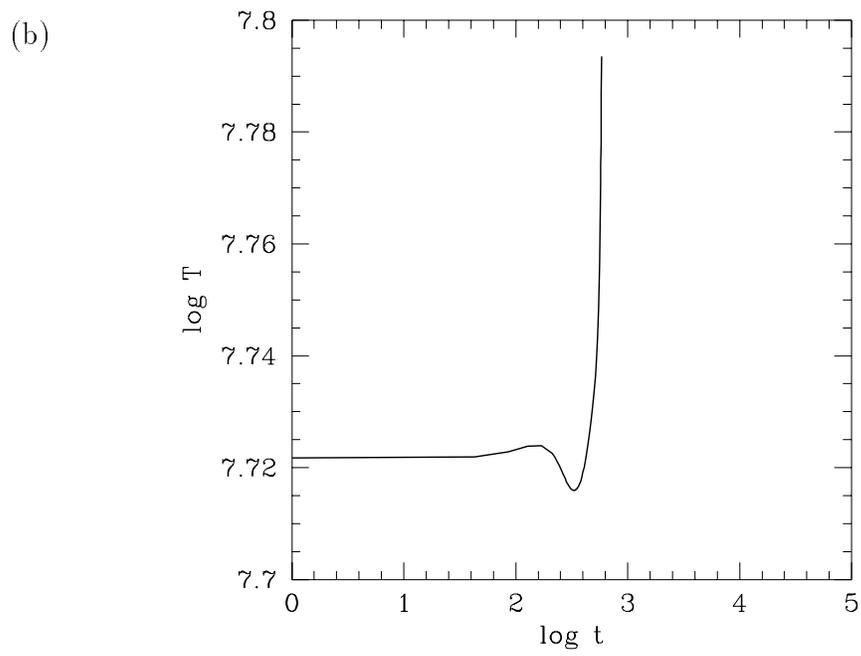
{\bf Figure 9:} (a) The evolutionary paths in the $\log T - \log \Sigma$
plane for $r=8\,r_{_G}$ for models with $\dot m = 0.09$ (solid line) and
0.1 (dashed line). The S-curve is shown by the dotted line. (b) The time
variation of the temperature at $r=8\,r_{_G}$ for the model with $\dot m =
0.09$. 

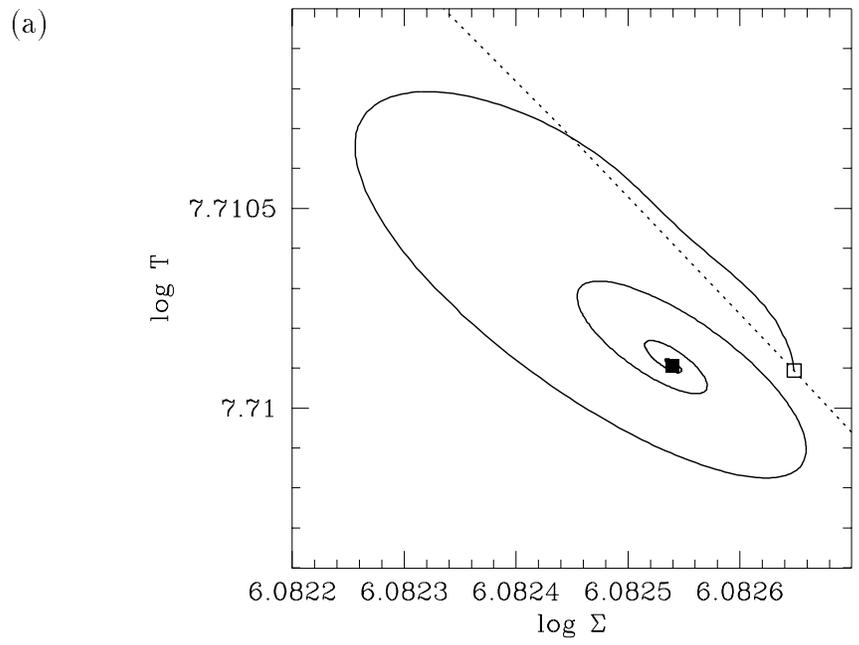
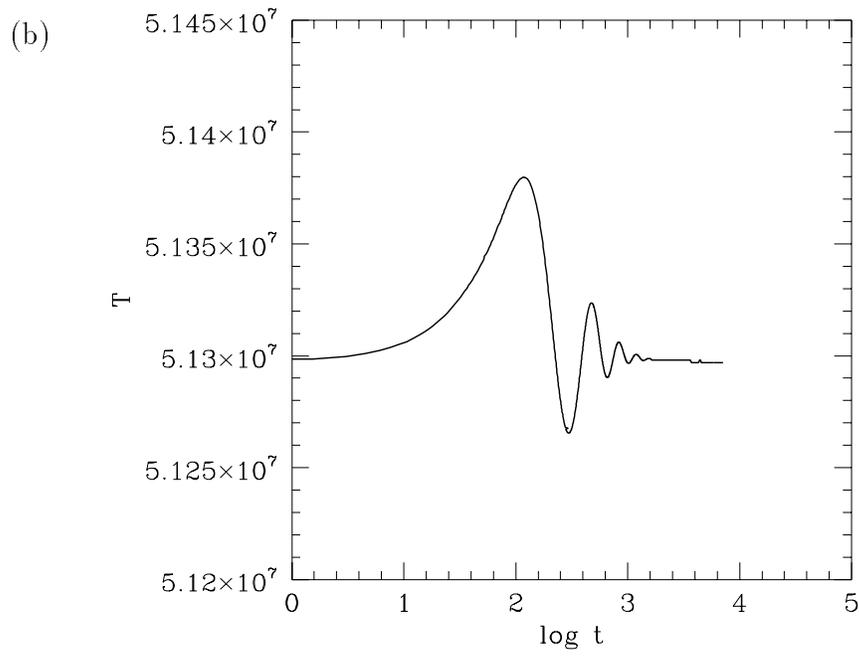
{\bf Figure 10:} (a) The evolutionary path in the $\log T - \log \Sigma$
plane for $r=8\,r_{_G}$ for the model with $\dot m = 0.08$. The S-curve is
shown by the dotted line. (b) The time variation of the temperature at
$r=8\,r_{_G}$ for the model with $\dot m = 0.08$. 
                 
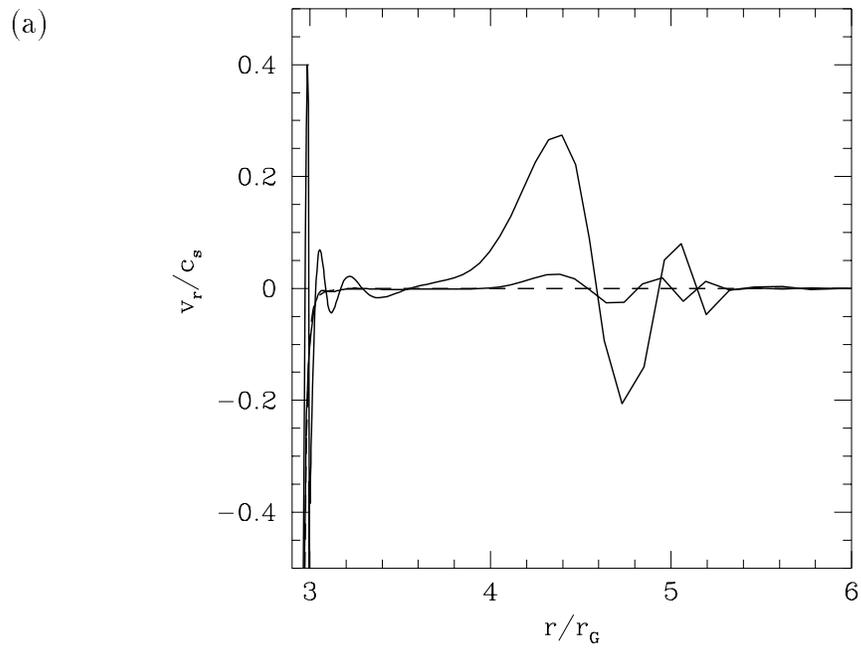
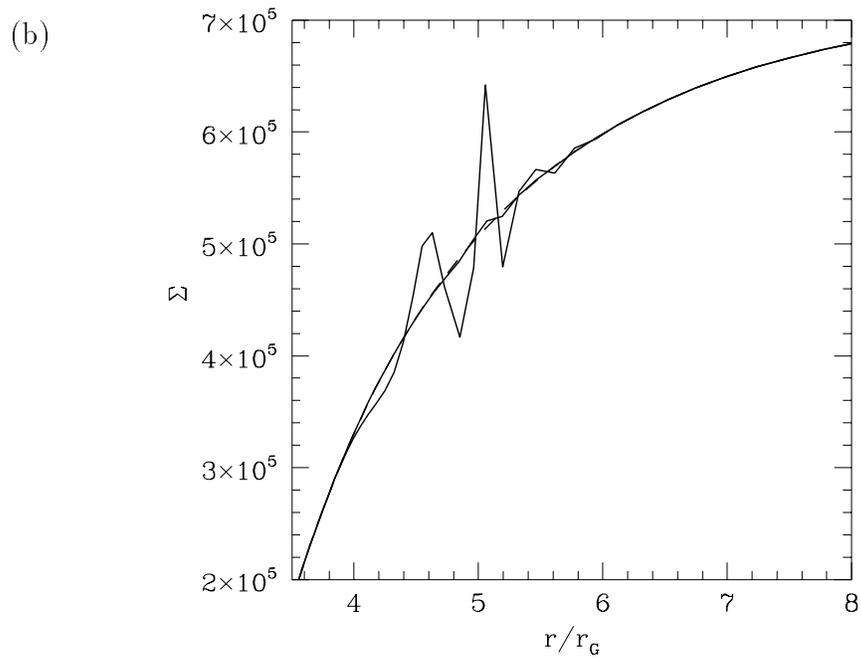
{\bf Figure 11:} The profiles of (a) Mach number and (b) surface density
for the model with $\dot m = 0.01$ at three different times: $t=0\,$s
(dashed line), $t = 76\,$s (lower amplitude perturbation) and $t =
87.7\,$s (higher amplitude perturbation). 

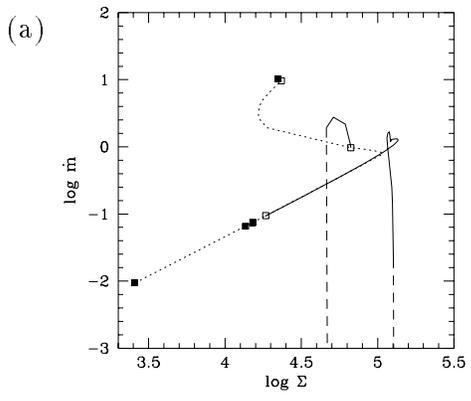
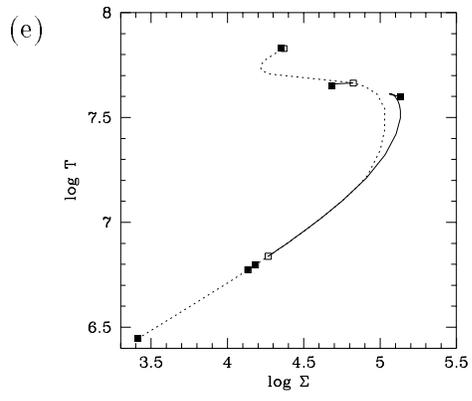
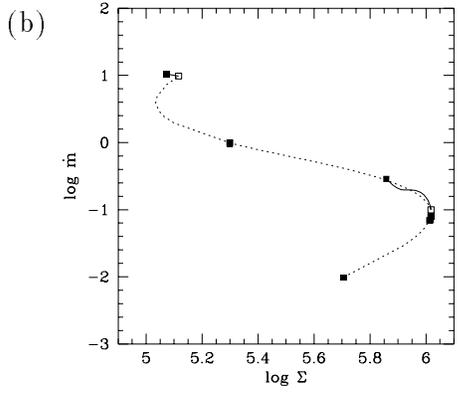
\end{document}